\documentclass[graybox]{svmult}

\usepackage{type1cm}        
\usepackage{longtable}
\usepackage{makeidx}         
\usepackage{graphicx}        
\usepackage{multicol}        
\usepackage[bottom]{footmisc}

\usepackage{newtxtext}       %
\usepackage{newtxmath}       

\makeindex             

\begin{document}

\title*{Conceptualizing Approaches to Critical Computing Education: Inquiry, Design and Reimagination }
\titlerunning{Conceptualizing Approaches to Critical Computing Education}
\author{Luis Morales-Navarro and Yasmin B. Kafai}
\institute{Luis Morales-Navarro \at University of Pennsylvania, Philadelphia, Pennsylvania, USA, \email{luismn@upenn.edu}
\and Yasmin B. Kafai \at University of Pennsylvania, Philadelphia, Pennsylvania, USA, \email{kafai@upenn.edu}}
%
%
\maketitle

\abstract*{As several critical issues in computing such as algorithmic bias, discriminatory practices, and techno-solutionism have become more visible, numerous efforts are being proposed to integrate criticality in K-16 computing education. Yet, how exactly these efforts address criticality and translate it into classroom practice is not clear. In this conceptual paper, we first historicize how current efforts in critical computing education draw on previous work which has promoted learner empowerment through critical analysis and production. We then identify three emergent approaches: (1) inquiry, (2) design and (3) reimagination that build on and expand these critical traditions in computing education. Finally, we discuss how these approaches highlight issues to be addressed and provide directions for further computing education research.}

\abstract{As several critical issues in computing such as algorithmic bias, discriminatory practices, and techno-solutionism have become more visible, numerous efforts are being proposed to integrate criticality in K-16 computing education. Yet, how exactly these efforts address criticality and translate it into classroom practice is not clear. In this conceptual paper, we first historicize how current efforts in critical computing education draw on previous work which has promoted learner empowerment through critical analysis and production. We then identify three emergent approaches: (1) inquiry, (2) design and (3) reimagination that build on and expand these critical traditions in computing education. Finally, we discuss how these approaches highlight issues to be addressed and provide directions for further computing education research.}

\section{Introduction}
\label{sec:1}
As CS education has gained an unprecedented momentum, becoming part of the K-12 curriculum over the past ten years and with increasing enrollment in higher education \cite{halvorson2020code}, several critical issues in computing as a discipline have become more visible: (1) algorithmic bias is pervasive, reinforcing historical inequities and damaging minoritized communities, (2) discrimination in industry against minoritized peoples in hiring and management practices continues, and (3) computing continues to be taught and learned as a value neutral subject \cite{gebru2020race}. Today we recognize that code is not neutral, computing technologies reflect the values and biases of their creators, and computing alone is not the solution to all problems, it often deepens existing ones, perpetuates existing concerns, or even causes new problems  \cite{costanza2020design}. But in most classrooms, computing is commonly introduced and experienced as a “value-neutral tool independent from society” (\cite{ko2020time}, p.31) without considering its societal and ethical implications \cite{vakil2018ethics, yadav2022toward}. Yet learning computing with disciplinary authenticity requires attending to its critical and political dimensions \cite{philip2021theories}.

More recently, efforts to foreground criticality in computing education have addressed this lack of concern for societal and ethical implications and limitations of the discipline through numerous proposals, among them: critical computational empowerment \cite{tissenbaum2017critical}, justice-centered efforts \cite{vakil2018ethics, costanza2020design, lin2021abstractions}, critical computational literacy \cite{lee2016none}, critical computing literacy \cite{scharber2021critical}, responsible computing \cite{Mozilla21}, computational action \cite{tissenbaum2019computational}, critical algorithmic literacy \cite{dasgupta2020designing}, abolitionist computing \cite{jones2021we}, computational empowerment \cite{iversen2018computational, dindler2020computational}, liberatory computing \cite{walker2022liberatory} and counter-hegemonic computing \cite{eglash2021counter}. While this proliferation of proposals highlights the growing importance given to critical issues in computing education, it is unclear what each approach means by criticality in computing, who is participating in these efforts, and how students and teachers are engaged in critical computing activities.

In this conceptual chapter, we take a step back from individual efforts by identifying common historical roots in human computer interaction (HCI) and in language arts and literacy (LAL) studies and how these perspectives foreshadowed today’s critical computing education (CCE) initiatives. We then identify and describe how three emergent approaches—(1) inquiry, (2) design, and (3) reimagination— address criticality in computing education. Finally, we discuss how these approaches highlight issues to be addressed and provide directions for future learning research and design.

\section{Historicizing criticality}
\label{sec:2}

Recent calls for considering the political dimensions of learning in design and research \cite{booker2014tensions, politics2017learning} have been followed by efforts in critical science literacy \cite{ryu2020refugee}, critical history and social studies \cite{jackson2020impact, hostetler2018unsilencing}, critical data literacies \cite{irgens2020data, stornaiuolo2020authoring}, and also in computer science education \cite{ko2020time, kafai2021revaluation}. Yet critical perspectives in computing and how computing is learned and taught have been present from the early days. For instance, Weizenbaum \cite{weizenbaum1976computer} argued that the technical innovations of computing did not necessarily promote social progress. Most importantly, he distinguished the differences in human and machine decision making, noting that computers lacked compassion in their calculations. From the education side, Papert \cite{papert1987information} pointed towards the fallacy of seeing the computer (or software) as the agent of change in student learning, considering people and culture as the driving forces of learning. Discussions in computing and education about criticality have historically developed in HCI around the Aarhus conferences community \cite{bertelsen2005critical} and around the critical literacies movement \cite{cazden1996pedagogy}. These two perspectives combined can be helpful to situate and historicize current efforts in CCE and better understand the development of different pedagogical approaches.

In language arts and literacy (LAL) studies critical literacies are seen as ways of being and doing through which learners participate in the world. Central here is Freire’s work on critical consciousness and literacy. Freire \cite{freire1970pedagogy} proposed an emancipatory model of literacy based on a dialectical relationship between humans and the world, where literacy is not seen as a collection of skills but rather a precondition for freedom, participation in the world and social empowerment within a wider project of social and political reconstruction. These ideas were further developed in the New London Group meetings where the multimodality of literacies and pedagogy were considered to propose that literacies must have a critical framing for learners to grow in their practice while consciously engaging with historical, social, and political contexts  \cite{cope1999introduction}. Today, critical literacies bring together ideas from marxist, queer, feminist, postcolonial, and critical race theories to inquire on the dynamics of power present in the world and the learning process. In a recent review, Vasquez and colleagues \cite{vasquez2019critical} argue that such a framework does not always involve taking a negative stance, but rather looking at issues from different perspectives, analyzing, suggesting and creating possibilities for change and improvement. The vast literature in critical literacies can be helpful in bringing CCE to classrooms. In fact, thinking about computing by applying frameworks from literacies is not new. The relationship between reading and writing code as extensions of literacies has been extensively discussed \cite{disessa2001changing, vee2017coding} by emphasizing how learning computing is necessary to fully participate in the world.

In HCI the term “critical computing” has a long, and often overlooked, history that dates back to 1975 when the first Aarhus conference convened to discuss the development of computing systems in context. These conversations at Aarhus emerged through dialogues on how computing could support workers with a particular interest in marxist approaches for understanding the design and use of technology in relation to class and power struggles \cite{nygaard1992many}. Over time, these discussions have evolved, expanding critical computing to “critical action, not only as workplace actionism, but also by integrating a broader scope of critical analysis and critical practice” (p. 3)—grounded in political, economic, and aesthetic theories—in how computing systems are designed and used in the workplace, education, and at home \cite{bertelsen2005critical}. The contributions of the Aarhus community and their perspectives on computing challenge many of the everyday practices of computing professionals and the ways we introduce learners to the field. For instance, within the Aarhus community, Burstall \cite{burstall1992computing} reflected that computing education that prioritizes technical concepts at the expense of critical awareness of other people and the environment loses the ability to accomplish the goals of the field.

\subsection{Empowerment}
\label{subsec:1}

Even though these critical perspectives in computing and education are products of different communities, contexts and conversations, taken together they provide useful underpinnings for CCE. Indeed the agency of learners, and the analysis as well as production of texts and code have been conceptualized as foundations for critical engagement in both traditions. To begin with, both HCI and LAL emphasize the need for empowerment or agency of learners. For instance, in Freire and Macedo’s \cite{freire1987literacy} emancipatory model of literacy, literacy is a precondition for empowerment, that is being able to fully participate in the world. Recently, critical literacies have been framed as a way of being and doing \cite{pandya2014moving}, extending literacies beyond an orientation for teaching and learning. This with the goal that learners can analyze and interrogate the micro features of texts and their macro—institutional, political, societal—conditions while focusing on how relations of power work, design and produce texts beyond classroom assignment \cite{vasquez2019critical}. The goal of literacy here is also one of empowerment, for learners to critically engage with the world to design social futures \cite{cazden1996pedagogy}. From the HCI perspective, Aarhus’ “critical computing” proposes that computing is a process of reality construction and transformation of the world \cite{floyd1992software}. This opens possibilities to reframe the development of computing applications from designing for requirements to designing for the opportunities of creating better, more just and equitable worlds \cite{floyd2002developing}. In this sense, novice learners must be empowered to use their curiosity and creativity to critically interrogate the history, implications and limitations of computing and create for the possibility of more equitable futures. How do these perspectives envision promoting agency or empowerment of learners? Two approaches stand out: one being reading or analysis, the other being writing or production of “text” or “code”.

\subsubsection{Analysis}
\label{subsubsec:1}
In both LAL and HCI analysis plays an important role in critical empowerment. For instance, in literacies, critical reading engages learners in reading beyond the words of the page. Indeed, for Freire “reading the world precedes reading the word” \cite{freire1987literacy} as learners can use critical literacy to make sense of their everyday lived experiences. Critical reading involves deconstructing texts (e.g., media, discourse, technologies) to question how these are constructed \cite{vasquez2019critical}. This kind of reading generates opportunities for “unpacking myths and distortions and building new ways of knowing and acting upon the world” (\cite{luke2014defining}, p. 22) by consciously investigating relationships of power, ideologies, and values in the process of reading \cite{cazden1996pedagogy}. As learners critique texts, they deconstruct and reconstruct them creating opportunities to disrupt, examine and sometimes dismantle problematic practices as well as to imagine alternatives, hypothesize how to change things and even take action \cite{vasquez2019critical}. Similarly, in the Aarhus community technological criticism also plays an important role for understanding the impact of computing and how computers are used. Here, reflection on the unconscious values embedded in computing applications and practices is central as well as investigating the ways in which technologies perpetuate oppression  \cite{sengers2005reflective}. 

\subsubsection{Production}
\label{subsubsec:2}
To promote agency, writing or production are important. The New London Group \cite{cazden1996pedagogy}, in reframing literacy studies, foregrounds the key role of designing new texts and redesigning existing texts for learners to participate in designing the future. This process of production requires that learners understand the positions from which they design so that they too read their own creations critically. This perspective resonates with Freire’s \cite{freire1985reading} idea that “reading the word is not only preceded by reading the world, but also by a certain form of writing it or rewriting” (p. 18). In this context, critical literacies are not only critical reading but also critical writing where writing means designing and redesigning texts in ways that aspire towards justice and can be socially transformative. 

Likewise, Christiane Floyd \cite{floyd2005}, who together with Kristen Nygaard played a key role in shaping the critical computing agenda within the Aarhus conferences, argues that in critical computing, understanding the impact of computing and how computers are used is not enough. She claims that critical computing is intertwined with the development of computing applications and the goal that these applications should have a positive impact in society. Furthermore, Floyd \cite{floyd2005} argues that there is a difference between being critical \textit{in} computing and being critical \textit{on} computing. Being critical \textit{in} computing is concerned with design and production of critical computing applications. This is distinct from being critical \textit{on} computing, which centers technological criticism over the design of applications that can address critical issues. CCE has the potential to engage learners in being critical \textit{on} computing through the analysis of computing applications discussed in the previous section. Yet, learners can also be critical \textit{in} computing by deeply engaging with the social and political implications of computing through the production of applications \cite{floyd2005}. From this perspective, social and technical factors are intertwined \cite{andelfinger2002intertwining} and the idea of computer programming must be widened to take into account the social context of software use and its development \cite{nygaard1986program}. Programming is framed as the means for non-technical ends or goals \cite{andelfinger2002intertwining} which requires investigating how the decisions made when coding have social and environmental implications \cite{nygaard1992many}.

\section{Approaches to Critical Computing Education}
Reviewing the historical roots of criticality in both computing and education provides a backdrop to examine how current research promotes CCE. From this review of HCI and LAL it emerges that CCE must empower learners to critically analyze computing, examine power and values, and interrogate the implications and limitations of computing applications. Likewise, CCE should include creating computing applications that aspire towards justice and that reflectively consider the limitations and implications of computing for people and the environment. In this section, we propose that out of these two traditions, three different pedagogical operationalizations or framings CCE emerge: (1) critical inquiry, (2) critical design, and (3) critical reimagination. Each of these approaches promotes the empowerment of learners through analysis (reading), production (writing), or combinations thereof (see Figure 1). Whereas our conceptualization of the inquiry approach stems out of the analysis (reading) tradition of critical literacy and the idea of being “critical on computing” from the Aarhus conferences, the design approach includes analysis as well as production or being “critical in computing.” Reimagination involves both analysis and production with the goal of envisioning more equitable speculative futures for computing. While these approaches are distinct, they are not mutually exclusive and at times may even be complimentary; in practice many CCE efforts draw on more than one approach.

\begin{figure}[h]
  \centering
  \includegraphics[width=\linewidth]{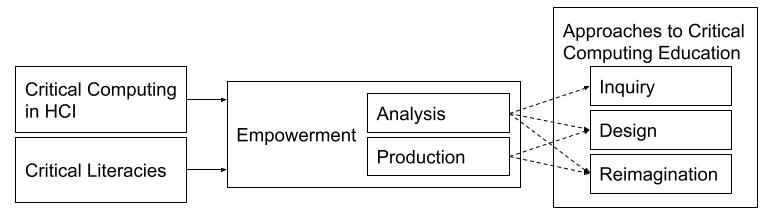}
  \caption{Relationships between historical underpinnings of criticality and approaches to critical computing education.}
\end{figure}

\subsection{Critical Inquiry}
One approach to CCE, which we call “critical inquiry” involves students inquiring on the implications of computing. For instance, in a social design experiment conducted by Vakil \cite{irgens2020data}, teenagers examined surveillance technologies in their communities and created infographics to explain the implications and limitations of these technologies. Through this experience, students explored the ethical tensions of how the surveillance technologies they encounter and use in their everyday lives not just reproduce but may amplify injustices in new ways, disproportionately targeting and impacting minoritized communities. In a discussing another iteration of the design experiment, Vakil and McKinney de Royston \cite{vakil2022youth} highlight how their program was designed to “foreground learning how to decode and unmake tech’s relationship with power through artistic, moral and humanistic inquiry” (p. 2) by positioning youth as philosophers of technology. Here youth, through the production of a documentary film, examined the ways in which computing technologies are used to surveil immigrants and also how immigrant communities resist surveillance through the use of technology. In a similar fashion, Vogel \cite{vogel2021programadores} discusses how bilingual secondary school students critiqued the educational technologies used at school by inquiring into their embedded values and particularly their raciolinguistic ideologies.  

Another example of critical inquiry is a course, co-constructed by Everson and colleagues with secondary students \cite{everson2022key}, in which learners explored inequities in access to computing education by creating data visualizations, investigated bias in machine learning, data privacy practices and their ethical implications. For instance, students researched how bias in machine learning models in everyday applications can have negative impacts in the lives of users and asked questions such as “What can we do?,” demonstrating a desire for change. Similarly, Walker and colleagues \cite{walker2022liberatory} propose a series of activities that can engage Black secondary students with what they call liberatory computing. Among these, they suggest students investigate data practices by analyzing how computing artifacts treat data and how youth produce data themselves and conduct “evocative audits” to research the positive and negative ways in which computing systems affect the socio-political realities of their communities.

Other efforts, at an undergraduate level, center on incorporating critical inquiry in introductory courses in which learners first encounter foundational technical concepts \cite{fisler2022approaches}. Lin \cite{lin2021abstractions}Lin [38], for example, argues that affordance analysis can be used in higher education to examine how algorithms and abstractions such as data structures have political implications. For instance, students could discuss how the use of binary trees in hiring decisions, autocomplete for search engines, and shortest paths for navigation systems can have problematic unintended consequences that reproduce systemic injustice. Furthermore, Lin \cite{lin2022cs} proposes Critical Comparative Data Structures and Algorithms as a justice-centered approach to teaching and learning in undergraduate computer science. This approach centers on critique of the values embedded in data structures and algorithm design and solutionist approaches to computing by constantly comparing dominant approaches to design justice approaches \cite{costanza2020design}. The goal here is for undergraduate students to engage beyond learning programming to understanding computing as a socio-technical field where the decisions they make while programming have implications. Lin highlights the importance of doing this work in a learner-centered environment where students can have agency to propose topics and examples for discussion that are relevant to their interests and lived experiences. Kirdani-Ryan and Ko \cite{kirdani2022house} provide another example of critical inquiry at an undergraduate level discussing how they embedded discussions and assignments about ethics and justice in  a low level software (computer systems) introductory course. In this course they framed computing as an object of critique and engaged students in investigating dominant and counter narratives in the discipline.

Whereas the examples discussed above incorporate critical inquiry throughout semester-long undergraduate courses, Horton and colleagues \cite{horton2022embedding} show that short isolated modules can also be engaging and informative for students. In their study, the inclusion of two 50-minute long modules during which students engaged in critical conversations around data privacy and ethics in introductory computing course (CS102) increased learner’s interest in ethical issues and self-efficacy in dealing with ethical issues. Having students be critical on computing in the same courses where they are learning technical concepts and skills can situate conversations and applications of technical concepts in their social and historical contexts.

The “critical inquiry” approach centers on critique—that is looking at issues from multiple perspectives, analyzing dynamics of power, and suggesting possibilities for change \cite{vasquez2019critical}, drawing on the tradition of LAL. This approach can be productive when integrated across introductory computing curricula in order to have learners consider how the decisions made while programming have social and environmental implications. That way, as novices learn technical concepts and programming skills these can be grounded in critical socio-technical conversations on computing \cite{nygaard1992many}. When engaging in critique, learning activities can connect to the lived experiences of students, like in Vakil’s experiment, by having them investigate the implications and limitations of the technologies they use or even further by critiquing their own creations. Critique can be empowering, giving students the rightful presence—that is legitimized membership and participation, where the community works towards justice by making injustice and social change visible \cite{calabrese2019designing}—to interrogate systems of oppression and question what computing is and how computing is enacted in the world. At the same time, learners can have agency to decide what issues they want to investigate. Yet CCE has the potential to go beyond inquiry by engaging learners in the production of critical computing applications.

\subsection{Critical Design}
A second approach to CCE, which we call “critical design”, centers on designing and redesigning with computing in ways that aspire towards justice and change. In this approach, learners critically design applications that address the needs of  communities. For instance, Lee and Soep \cite{lee2016none} argue that when young people design applications to address their own community problems they create opportunities for themselves and their communities, and that “only through production of these digital tools will youth develop the agency required to make the changes they want to see” (p. 481). Through their experiences at Youth Radio they have documented how teenagers create mobile and web applications for their communities, including, for example, an app called “Know Your Queer Rights” that provided resources to support LGBTQ+ youth \cite{soep2021} or an app to track and raise awareness of gentrification \cite{lee2016none}. These examples show how young people can take action to address inequities in their communities and design with computing.

Other approaches to critical design have students design for communities other than their own. Bar-El and Worsley \cite{bar2021making} discuss a university course for undergraduate and graduate students centered around accessibility, designed to encourage students to explore computing as a field that promotes democratization by having learners design tools and activities to broaden participation in CS. In this intervention, students were embedded in the spaces for which they designed and were in dialogue with community members. The authors argue that by creating real-life projects for real communities, learners can benefit from interrogating the history and assumptions of computing and its applications \cite{bar2021making}. Since the majority of participants in the course were engineering students, most already had a background in computing which was helpful for students to develop complex projects that included, for example, browser extensions to audit web accessibility or multimodal interfaces for design tools to support blind people engaging with computer aided design.

Tissenbaum and colleagues \cite{tissenbaum2017critical} propose that empowerment comes from creating interventions in the real and physical world in which learners analyze the problems of their communities and create real solutions. For them, learners must have opportunities to create applications that have an impact on their communities from the moment they begin to learn how to code \cite{tissenbaum2019computational}. This contrasts with common approaches to computing education that prioritize learning technical skills first and working on real-world applications later. In this line, Van Wart and colleagues  \cite{van2014apps} present a case study of two high school students participating in a 12-week App design program where students investigated community problems, identified local needs and designed mobile applications to address these needs. They argue that in this “practice-oriented computer science learning environment” students accessed disciplinary concepts and skills and demonstrated their competency \textit{while} working on projects that were relevant to their communities and addressed issues of justice. Here learners had agency to learn the concepts that were most relevant to their goals in an environment that recognized them as local experts using computing as a tool to address real world issues.

Learners can also engage with critical design by making small classroom projects that address critical issues related to the technologies they use everyday. Vogel \cite{vogel2021programadores}, for instance, presents a compelling case study of how in a secondary school classroom after the teacher introduced voice recognition functions in a programming environment, bilingual youth worked on designing voice-based interfaces that could understand their translanguaging practices. 

While engaging in critical design, learners have the opportunity to participate in critiquing and producing computing applications. Critical design aligns well with the ideas from the Aarhus conference particularly seeing programming as a social activity where writing code is the means for non-technical ends or goals \cite{andelfinger2002intertwining}. It also addresses coding, in a similar fashion to how critical literacies theorize writing, as the process in which learners design and redesign in ways that aspire towards justice while reading their own creations critically \cite{vasquez2019critical}. Critical design is also an opportunity to design computing applications beyond requirements, thinking about the opportunities \cite{floyd2002developing} that computing gives creators and users to make more equitable worlds and to inquire into the sociopolitical values and purposes of making with computating technologies \cite{vossoughi2016making}. Furthermore, analyzing problems in their communities and designing computing applications to address these problems can support the development of learners’ rightful presence, empowering students to be and do critical computing in authentic and meaningful ways. But empowering students also means giving them the creative power and freedom to pursue their own interests and projects. These design activities can connect to students’ lives beyond deficit views of their experiences \cite{jones2021we} recognizing how the cultural wealth of their communities can have a positive impact in the learning environment and the computing applications learners create. This requires thinking about cultural context beyond making superficial connections to computing concepts so that learners can participate in their cultures while computing, draw from their own culture and even remake it by investigating authentic problems and concerns they can engage with in their designs \cite{blikstein2008travels, hooper2008looking}. 

\subsection{Critical Reimagination}
A third approach which we call “critical reimagination” involves rethinking the present and the past to critically reimagine computing and technology and create more equitable and just futures. Holbert and colleagues \cite{holbert2020afrofuturism}, for instance, created the "Remixing Wakanda'' project in which Black female teenagers used their personal stories to design speculative artifacts that address social and environmental injustice. Bringing together Afrofuturist and constructionist ideas, participants imagined and created possible futures where their communities thrived and where technology helped create a more equitable and sustainable world. They reimagined how technology could be used proactively to improve their communities. They created speculative computing projects such as a cloak that expressed information about the health, wellness, mood and identity of its wearer or a trash can that converted waste into energy. Other efforts have focused on having students reimagine computing cultures that are equitable where female, Black and brown youth can fully participate. Efforts employing restorying \cite{shaw2021promoting} build on Black feminist perspectives for youth to imagine and build the worlds and computing cultures they want to live and participate in. Shaw and colleagues \cite{shaw2021promoting} describe how youth engaged in interrogating dominant narratives about CS and crafted computational artifacts that reimagined CS, its values, who can participate and how it is done. In contrast to the ‘Remixing Wakanda’ effort, which drew on the futuristic visions promoted in a popular movie, the restorying effort drew on the often forgotten historical connections between computing and textile work from Jacquard’s loom as a predecessor to modern computing to the use of quilts by Black women to address social issues.

Speculating about the future of computing at a university level can often involve engaging in “Black Mirror” exercises drawing on a popular science fiction TV series to inquire into the possible ethical dilemmas and social impacts of computing applications \cite{klassen2022run}. These are common exercises in undergraduate computing courses on ethics and society. While here learners engage in being critical about computing and imagining futures, these are often dystopian. Yet, as Klassen and Fiesler \cite{klassen2022run} suggest these activities have the potential to also be suitable to imagine more ethical and just futures for computing.

Critical reimagination engages learners with the social and political reconstruction discussed by critical literacy scholars such as Giroux \cite{giroux1988literacy} that requires critique, production and developing an empowered voice to imagine and create alternatives for liberation and justice. As hooks \cite{hooks1991theory} argues, imagination is emancipatory because thinking of possible futures involves critical thinking, analysis, and reflection of the present as well as a desire to build better worlds. In this approach when learners engage in production they design beyond requirements for the opportunity of creating better worlds. Learners address critical computing as emancipation  \cite{dearden2005choosing} and through their speculative computational artifacts transform the world by constructing and proposing new realities \cite{floyd1992software}.They are empowered to reimagine what can be done with computing and what computing can be, having the rightful presence to question and rethink the discipline. Yet in critical reimagination activities it is important to provide space for student agency so that learners develop their own voices while accomplishing their desired computational goals. Reimagination connects the past, present and future, understanding that technology is not created in a void but builds on previous traditions. With critical reimagination the future is not pre-established but rather learners engage in what Freire \cite{freire1970pedagogy} calls hopeful “revolutionary futurity,” looking at the past to understand themselves and addressing how the present must change in order to imagine and build the future.

\section{Considerations for the Learning Design and Research}
\label{sec:4}

By historicizing how criticality has been approached in computing and education we situate CCE in both fields. The three approaches we discussed as framings for learning and instruction emerged from our analysis of efforts to design and research criticality in computing. From our analysis, we see that some of these current efforts in CCE align with the three proposed framings while others engage with more than one (see Table I.). For instance, some justice efforts \cite{lin2021abstractions} and liberatory computing \cite{walker2022liberatory} align with critical inquiry while critical computational empowerment \cite{tissenbaum2017critical}, critical computational literacy \cite{lee2016none} and computational action \cite{tissenbaum2019computational} prioritize critical design. Abolitionist computing \cite{jones2021we} and counter-hegemonic computing \cite{eglash2021counter},tend to align better with critical inquiry and reimagination. Other justice-centered efforts \cite{vakil2018ethics}, responsible computing \cite{Mozilla21}, critical computing literacy \cite{scharber2021critical}, and critical algorithmic literacy \cite{dasgupta2020designing} engage both critical inquiry and design. Computational empowerment \cite{iversen2018computational, dindler2020computational} and justice-centered efforts such as Costanza-Chock’s design justice \cite{costanza2020design} address critical inquiry, design and reimagination.

\begin{longtable}{l|l}
\caption{Different CCE efforts and their engagement with criticality.} \\ \textbf{CCE Effort} & \textbf{Engagement with Criticality}\\ \hline
\endfirsthead
\endhead
\begin{tabular}[c]{@{}l@{}}\textbf{Abolitionist Computing} \cite{jones2021we}\\ Integrating an abolitionist framework to \\ CS to open up world-building possibilities \\ that affirm Black Life.\end{tabular} & \begin{tabular}[c]{@{}l@{}}\textit{Inquiry:} Examining anti-Blackness\\ in CS and CS education.\\ \textit{Reimagination}: Reimagining CS\\ through Black Life-affirming \\ world-building projects.\end{tabular} \\ \hline
\begin{tabular}[c]{@{}l@{}}\textbf{Computational Action} \cite{tissenbaum2019computational}\\ Engaging youth take action with \\ computing by making applications that \\ have an impact in their communities.\end{tabular} & \begin{tabular}[c]{@{}l@{}}\textit{Design:} Students research problems\\ in their communities and learn\\ computing by creating applications\\ to address these problems.\end{tabular} \\ \hline
\begin{tabular}[c]{@{}l@{}}\textbf{Computational Empowerment} \cite{dindler2020computational, iversen2018computational}\\ Engaging students in understanding\\ computing technologies and their effects\\ on their lives and society by critically \\ constructing and deconstructing \\ computing artifacts.\end{tabular} & \begin{tabular}[c]{@{}l@{}}\textit{Inquiry:} Students critique and assess \\ everyday technologies by \\ considering their impact and \\ implications.\\ \textit{Design:} Students create computing \\ applications that address problem \\ situations.\\ \textit{Reimagination:} Students co-create \\ the future of computing by critically\\ decoding and coding artifacts.\end{tabular}\\ \hline
\begin{tabular}[c]{@{}l@{}}\textbf{Counter-hegemonic Computing} \cite{eglash2021counter}\\ Engaging with Black students’ \\ counter-hegemonic practices \\in computing.\end{tabular}                                                                                                                              & \begin{tabular}[c]{@{}l@{}}\textit{Inquiry:} Students examine negative\\ and positive frames of reference in \\ computing by considering power \\ and identity at societal and\\ individual levels.\\\textit{Reimagination:} Using computing for \\ emancipating counter-hegemonic\\ practices.\end{tabular}                                                                                                                                                                                  \\ \hline
\begin{tabular}[c]{@{}l@{}}\textbf{Critical Algorithmic Literacies} \cite{dasgupta2020designing}\\ Enabling youth to critique and understand\\  the algorithmic systems they encounter\\  everyday.\end{tabular}                                                                                                      & \begin{tabular}[c]{@{}l@{}}\textit{Inquiry:} Children analyze data sets \\ related to the world in which they \\ live. \\ \textit{Design:} Children create and \\ experiment with algorithms within \\ “sandboxes for dangerous ideas.”\end{tabular}                                                                                                                                                                                                                                         \\ \hline
\begin{tabular}[c]{@{}l@{}}\textbf{Critical Computational Empowerment} \\\cite{tissenbaum2017critical}\\ Engaging young learners in creating\\ personally meaningful applications that \\ have impact in the real world\end{tabular}                                                                                     & \begin{tabular}[c]{@{}l@{}}\textit{Design:} Students research problems\\ in their communities and learn \\ computing by creating applications\\ to address these problems.\end{tabular}                                                                                                                                                                                                                                                                                                  \\ \hline
\begin{tabular}[c]{@{}l@{}}\textbf{Critical Computational Literacy} \cite{lee2016none}\\ Engaging young people in creating \\ projects that address inequities and \\ injustice while learning to code.\end{tabular}                                                                                & \begin{tabular}[c]{@{}l@{}}\textit{Design:} Students research problems\\ in their communities and learn\\ computing by creating applications\\ to address these problems.\end{tabular}                                                                                                                                                                                                                                                                                                   \\ \hline
\begin{tabular}[c]{@{}l@{}}\textbf{Critical Computing Literacy} \cite{scharber2021critical}\\ Centering doing and being with\\ computing to broaden participation of\\ girls in the discipline.\end{tabular}                                                                                                          & \begin{tabular}[c]{@{}l@{}}\textit{Design:} Youth create applications\\ to address needs in their communities.\end{tabular}                                                                                                                                                                                                                                                                                                                                                              \\ \hline
\begin{tabular}[c]{@{}l@{}}\textbf{Design Justice}\cite{costanza2020design}\\ Remaining a community-centered \\ alternative to computing where\\ marginalized communities inquire on \\ the implications of technology to \\explicitly challenge structural \\inequities through design.\end{tabular} & \begin{tabular}[c]{@{}l@{}}\textit{Inquiry:} Communities research the\\ implications of computing \\ applications to propose alternatives \\ that foster justice.\\ \textit{Design:} Communities design \\ computing applications and tools to \\ create a more equitable and \\ sustainable world.\\ \textit{Reimagination:} Transform \\ computing through a \\ community-centered design \\ approach that aspires towards \\ collective liberation and ecological \\ sustainability.\end{tabular} \\ \hline
\begin{tabular}[c]{@{}l@{}}J\textbf{ustice-centered Approaches to CS} \cite{lin2022cs}\\ Constantly comparing dominant to CS\\ approaches to design justice approaches\\ particularly with regards to the values \\ embedded in data structures and \\ algorithms.\end{tabular}                           & \begin{tabular}[c]{@{}l@{}}\textit{Inquiry:} Engage learners in\\ investigating the values and\\ implications of technical decisions\\ made while programming.\end{tabular}                                                                                                                                                                                                                                                                                                              \\ \hline
\begin{tabular}[c]{@{}l@{}}\textbf{Justice-centered Computing} \cite{vakil2018ethics}\\ Interrogating the sociopolitical context \\of CS education through critical inquiry \\on the curriculum, design of learning\\ environments and purpose of CS\\ education.\end{tabular}                                & \begin{tabular}[c]{@{}l@{}}\textit{Inquiry:} Students research ethics, \\ power and politics of computing\\ technologies they use everyday. \\\textit{Design:} Students learn computing\\ with a purpose, investigating \\ community problems and designing \\ applications to address them.\end{tabular}                                                                                                                                                                                     \\ \hline
\begin{tabular}[c]{@{}l@{}}\textbf{Liberatory Computing} \cite{walker2022liberatory}\\ Computing curriculum to motivate \\and prepare Black students to address \\racism embedded in society.\end{tabular}                                                                                           & \begin{tabular}[c]{@{}l@{}}\textit{Inquiry:} Students research issues of\\ racism and inequity in computing \\ with the goal of becoming data\\ activists.\end{tabular}                                                                                                                                                                                                                                                                                                                  \\ \hline
\begin{tabular}[c]{@{}l@{}}\textbf{Responsible Computing} \cite{Mozilla21}\\ Approaching computing as a\\ socio-technical field while students \\ design artifacts.\end{tabular}                                                                                                                           & \begin{tabular}[c]{@{}l@{}}\textit{Inquiry:} Students discuss issues of\\ justice, implications and\\ consequences of computing.\\ \textit{Design:} Students design computing\\ artifacts that take into account \\ societal implications.\end{tabular} \\ \hline
\end{longtable}

In identifying the different directions criticality assumes in each of the framings, we also noticed several issues that require special attention in designing and researching learning tools, activities, and environments: (1) who is involved in addressing critical issues in computing, (2) how to avoid the pitfalls of techno-solutionism, (3) how learners engage creatively with criticality, and (4) how do we connect learning disciplinary skills and concepts with criticality, and (5) how to support teachers in bringing CCE to their classrooms? 

First, criticality in computing should be addressed by all. In reviewing the various efforts in critical inquiry, design and reimagination we noticed that these involved mostly students from historically excluded communities. While this was probably done with the best intentions, to broaden participation, we should also be aware that such efforts could put a double “burden” on learners of minoritized communities \cite{jones2021we}, first as those who predominantly experience marginalization and discrimination when using technologies and participating in computing, and then as those tasked to identify challenges, develop new solutions, fix problems and reimagine alternate futures. This is important work but it is essential that all learners are engaged in CCE.

Second, while having students create computing applications that address real world problems is beneficial, it is equally important to consider what already exists in the communities that learners engage with, encourage them to be critical towards their own work and avoid the pitfalls of techno-chauvinism and techno- solutionism—the beliefs that technology is always the solution \cite{costanza2020design}. Indeed, the critical design and reimagination approaches must acknowledge the limitations of computational solutions, questioning the possible implications of using computing to “fix the world.” 

Furthermore, we must consider how learners engage creatively with criticality. In inquiry, design and reimagination learners can connect to their personal interests and their lived experiences, to think, create and share their ideas with their peers. To do this, learners’ agency must be authentic, giving space for students to decide what issues to investigate and what kind of artifacts they want to create. This with the goal that they can engage in what Freire \cite{freire1970pedagogy} calls the “creative transformation” of understanding, coding and participating in the world. Ultimately, while we want learners to “have the opportunity to experience the full conceptual and expressive powers of coding” (\cite{resnick2020coding}, p. 121), we also see a need to connect criticality back to creativity and curiosity rather than to consider creativity as a distinct engagement with code. Several of the examples we presented in different framings of CCE provide compelling illustrations how creative engagement can be coupled with critical inquiry or design.

We should also consider how we join criticality with efforts for developing technical coding skills rather than seeing them as separate enterprises. In K-12, we take note that efforts in CCE predominantly take place in out-of-school settings. This is partly because current curricular frameworks and standards are very much focused on teaching and learning of computational skills and concepts, often relegating criticality to the sidelines. Yet alternatives are possible. In Denmark, for instance, the national K-12 computing curriculum framework \cite{caspersen2022informatics} inherited many of the ideas of the Aarhus conferences \cite{iversen2018computational, dindler2020computational} giving computational empowerment the same importance as computational thinking. Here empowerment is approached as the concern for learners to understand how computing affects their lives and society to creatively and critically participate in the construction of computing applications. On the other hand, recent efforts in higher education \cite{fisler2022approaches, lin2022cs, kirdani2022house} push for the integration of critical inquiry in introductory courses. There is plenty of potential to further critical design and reimagination in higher education, particularly in non-introductory courses, where students' existing technical knowledge can lead to novel applications as illustrated in the work of Bar-El and Worsley \cite{bar2021making}.

Finally, supporting teachers to integrate CCE in their classrooms is crucial. Most of the examples discussed in this chapter are from researcher-led, small interventions. Yet, scaling CCE education may be challenging as teachers are often trained to teach computing from a technical perspective only and as a value neutral subject \cite{ko2020time}. While recent efforts, such as the publication of a text-book for secondary teachers \cite{ko2021critically}, a graphic novel that address issues of critical computing \cite{ryoo2022Power}, or a site with crowd-sourced critical coding activities \cite{CriticalCode} may support teachers in learning about and engaging their students with CCE, it is also crucial to partner with educators to co-design professional development experiences. Researchers and teachers should also partner in co-designing and redesigning classroom activities and curriculum to engage with criticality. For instance Jayathirtha and colleagues \cite{jayathirtha2021redesigning} illustrate how existing coding activities can be re-configured so that students consider implications and limitations in human-computer interaction.

\section{Conclusions}
\label{sec:5}

Our goal in this conceptual paper is not to pit approaches against one another but to recognize that inquiry, design and reimagination each offer valuable ways to create learning activities that can engage with criticality in computing education research. Indeed, as Vakil and Higgs \cite{vakil2019s} write, the goal is that learners “can understand, analyze, critique, and reimagine the technologies that shape everyday lives” (p.31). The approaches, in fact, offer partial but complimentary ways for addressing criticality. Inquiry focuses on understanding the underpinning of technologies and power structures. Design promotes understanding criticality while making computing applications, creating a relevant context for learning technical skills and concepts, and reimagination centers on rethinking the present and the past to critically reimagine computing for the future. We hope our proposed framings may serve as tools for learning scientists to further design, investigate, and harness their potential in driving disciplinary critical engagement not only in computing but also in other fields.

In conceptualizing the proposed operationalizations of CCE, we are reminded that being critical involves both understanding the role of computing in society and of society in computing \cite{kafai2021revaluation}. This requires not only discussing critical issues but deeply inquiring, designing and reimagining how computing can be transformed to be more just and equitable and to positively impact the lives of communities. Ultimately, as we saw in our historization of CCE, the goal is to empower learners to fully participate in computing through critique and production. We hope that our conceptualization of three emerging operationalizations contributes to advance our understanding of criticality in learning computer science.

\section*{Acknowledgements}
An earlier version of this chapter titled “Conceptualizing three approaches for integrating criticality in K-12 computing education” was presented and published in the proceedings of the International Conference of the Learning Sciences (ICLS) 2022.

%
%
\bibliographystyle{spmpsci.bst}
\bibliography{sample-basex}
%

%
%


\end{document}